\documentclass{article}
\usepackage{amsfonts}
\usepackage{amsmath}

\input{tcilatex}

\begin{document}

\title{Properties of $2\times 2$ $h$-deformed quantum (super)matrices\thanks{%
The project supported by National Natural Science Foundation of China under
Grant No. 10075042}}
\author{Yun Li\thanks{%
Email: yunlee@mail.ustc.edu.cn} and Sicong Jing \\
\textit{Department of Modern Physics,}\\
\textit{University of Science and Technology of China, }\\
\textit{Hefei, Anhui 230026, P.R.China} }
\maketitle

\begin{abstract}
We investigate the $h$-deformed quantum (super)group of $2\times 2$ matrices
and use a kind of contraction procedure to prove that the $n$-th power of
this deformed quantum (super)matrix is quantum (super)matrix with the
deformation parameter $nh$.

\textit{keywords}: $h$-deformed, quantum group, quantum matrix
\end{abstract}

There are two distinct quantizations for general Lie group. One is the
well-known Drinfeld-Jimbo $q$-deformed quantum group and the other is the
so-called Jordanian $h$-deformed one. For group $GL(2)$ there exist only
such two deformation (up to isomorphism) with a central quantum determinant: 
$GL_{q}(2)$ and $GL_{h}(2)$\cite{1}. One can obtain these quantum groups by
deforming the coordinates of a linear plane to be noncommutative objects. In
detail, the deformed quantum groups act on the $q$-plane, with the relation $%
xy=qyx$ and the $h$-plane, with the relation $xy-yx=hy^{2}$, respectively.
The $q$-deformed quantum group has been intensively studied\cite{2} because
it is closely related with the solution of Yang-Baxter equation and the
theory of braids. In \cite{3}, it is found that the deformation parameter of 
$h$-deformed quantum group is naturally dimensional (like the known $k$%
-deformation). This is quite intriguing for possible physical applications.
Recently geometrical structure of the $h$-plane has been thoroughly discussed%
\cite{4} and quantum mechanics based on the plane was constructed\cite{5}.
These work are helpful to understand $h$-deformed quantum group.

In \cite{6}, $q$-deformed quantum group $GL_{q}(2)$ was investigated and it
was shown that the $n$-th power of a $q$-deformed quantum matrix corresponds
to the $n$-th power of the deformation parameter $q$. Also the same property
of $q$-deformed quantum supergroup $GL_{q}(1\mid 1)$ was discussed in \cite%
{7}. And in \cite{1}, it was pointed out that the $n$-th power of a $h$%
-deformed quantum matrix is also quantum matrix with the deformation
parameter $nh$. In \cite{8}, it was found that the $h$-deformed quantum
group can be obtained from the $q$-deformed quantum Lie group through a
singular limit $q\rightarrow 1$ of a linear transformation. This method is
called as the contraction procedure.

In this paper we use the contraction method to obtain the same result
pointed out in \cite{1}. We also extend the discussion to deformed quantum
supergroup and show that the $n$-th power of a $h$-deformed quantum
supermatrix is also quantum supermatrix with the deformation parameter $nh$.
Not only do we use the sigular linear transformation between the two kinds
of quantum (super)groups to obtain the results, but also we use the
similarity transformation between the corresponding $R$-matrices for
different quantum groups to obtain them.

To begin with we define the $q$-deformed quantum matrix as

\begin{equation}
M^{^{\prime }}=%
\begin{pmatrix}
a^{^{\prime }} & b^{^{\prime }} \\ 
c^{^{\prime }} & d^{^{\prime }}%
\end{pmatrix}%
\text{ .}  \label{1}
\end{equation}%
Throughout this paper we denote $q$-deformed objects by primed quantities.
The unprimed ones represent tansformed deformed objects. The Manin quantum
plane satisfying the commutation relation $xy=qyx$ can be considered
covariant with respect to the action of the quantum group $GL_{q}(2)$. $%
M^{^{\prime }}\in GL_{q}(2)$ means the following commutation relations are
fulfiled

\begin{eqnarray}
a^{^{\prime }}c^{^{\prime }} &=&qc^{^{\prime }}a^{^{\prime }},\text{ \ }%
b^{^{\prime }}d^{^{\prime }}=qd^{^{\prime }}b^{^{\prime }},\text{ \ }\left[
a^{^{\prime }},d^{^{\prime }}\right] =\left( q-q^{-1}\right) b^{^{\prime
}}c^{^{\prime }},  \label{2} \\
a^{^{\prime }}b^{^{\prime }} &=&qb^{^{\prime }}a^{^{\prime }},\text{ \ }%
c^{^{\prime }}d^{^{\prime }}=qd^{^{\prime }}c^{^{\prime }},\text{ \ }%
c^{^{\prime }}b^{^{\prime }}=b^{^{\prime }}c^{^{\prime }}\text{ .}  \notag
\end{eqnarray}%
Here $\left[ ,\right] $ stands for the commutator. The quantum determinant $%
D^{^{\prime }}$ which is defined as

\begin{equation}
D^{^{\prime }}=\det_{q}M^{^{\prime }}\equiv a^{^{\prime }}d^{^{\prime
}}-qc^{^{\prime }}b^{^{\prime }}  \label{3}
\end{equation}%
is central. One can obtain $h$-deformed quantum group $GL_{h}(2)$ with the
following transformation

\begin{equation}
M^{^{\prime }}=gMg^{-1}  \label{4}
\end{equation}%
where

\begin{equation}
g=\left( 
\begin{array}{cc}
1 & \frac{h}{q-1} \\ 
0 & 1%
\end{array}%
\right) \text{ .}  \label{5}
\end{equation}%
A simple calculation shows that the $q$-deformed matrix elements and the
corresponding tansformed ones fulfil the following relations:

\begin{eqnarray}
a^{^{\prime }} &=&a+\frac{h}{q-1}c,\text{ \ \ }b^{^{\prime }}=b+\frac{h}{q-1}%
(d-a)-\frac{h^{2}}{(q-1)^{2}}c,  \label{6} \\
c^{^{\prime }} &=&c,\text{ \ \ \ \ \ \ \ \ \ \ \ \ \ }d^{^{\prime }}=d-\frac{%
h}{q-1}c\text{ .}  \notag
\end{eqnarray}%
Substituting the above relations into the $q$-commutation relations$(2)$ and
taking the limit of $q\rightarrow 1$, one can find that the entries of the
transformed quantum matrix $M$ satisfy the following commutation relations
of the $GL_{h}(2)$

\begin{eqnarray}
\left[ a,c\right] &=&hc^{2},\text{ \ \ }\left[ d,b\right] =h(D_{h}-d^{2}),%
\text{ \ }\left[ a,d\right] =hdc-hac,  \label{7} \\
\left[ d,c\right] &=&hc^{2},\text{ \ \ }\left[ b,c\right] =hac+hcd,\text{ \ }%
\left[ b,a\right] =h(a^{2}-D_{h})\text{ .}  \notag
\end{eqnarray}%
Here the corresponding $h$-deformed determinant is

\begin{equation}
D_{h}=\det_{h}M\equiv ad-cb-hcd  \label{8}
\end{equation}%
which is also central. The $h$-deformed quantum plane satisfying the
relation $xy=yx+hy^{2}$ can also be considered covariant with respect to the
action of the above quantum group $GL_{h}(2)$.

For $n$-th power of $M^{^{\prime }}$, from $(4)$, we obtain

\begin{eqnarray}
M^{^{\prime }n} &=&(gMg^{-1})^{n}  \label{9} \\
&=&gM^{n}g^{-1}\text{ .}  \notag
\end{eqnarray}%
If we denote $M^{^{\prime }n}$, $M^{n}$ as$\left( 
\begin{array}{cc}
a_{n}^{^{\prime }} & b_{n}^{^{\prime }} \\ 
c_{n}^{^{\prime }} & d_{n}^{^{\prime }}%
\end{array}%
\right) $ and $\left( 
\begin{array}{cc}
a_{n} & b_{n} \\ 
c_{n} & d_{n}%
\end{array}%
\right) $, respectively, the relations between them can be obtained in a
similar way to get Eq.(6). They are given by

\begin{eqnarray}
a_{n}^{^{\prime }} &=&a_{n}+\frac{h}{q-1}c_{n},\text{ \ \ }b_{n}^{^{\prime
}}=b_{n}+\frac{h}{q-1}(d_{n}-a_{n})-\frac{h^{2}}{(q-1)^{2}}c_{n},  \notag \\
c_{n}^{^{\prime }} &=&c_{n},\text{ \ \ \ \ \ \ \ \ \ \ \ \ \ }%
d_{n}^{^{\prime }}=d_{n}-\frac{h}{q-1}c_{n}\text{ .}  \label{10}
\end{eqnarray}%
But because $M^{^{\prime }n}$ belongs to $GL_{q^{n}}(2)$\cite{6}, the matrix
elements of $M^{^{\prime }n}$ will satisfy the commutation relations$\left(
2\right) $ only with the deformed parameter $q$ replaced by $q^{n}$. So we
can have the following relations

\begin{eqnarray}
a_{n}^{^{\prime }}c_{n}^{^{\prime }} &=&q^{n}c_{n}^{^{\prime
}}a_{n}^{^{\prime }},\text{ \ }b_{n}^{^{\prime }}d_{n}^{^{\prime
}}=q^{n}d_{n}^{^{\prime }}b_{n}^{^{\prime }},\text{ \ }\left[
a_{n}^{^{\prime }},d_{n}^{^{\prime }}\right] =\left( q^{n}-q^{-n}\right)
b_{n}^{^{\prime }}c_{n}^{^{\prime }},  \notag \\
a_{n}^{^{\prime }}b_{n}^{^{\prime }} &=&q^{n}b_{n}^{^{\prime
}}a_{n}^{^{\prime }},\text{ \ }c_{n}^{^{\prime }}d_{n}^{^{\prime
}}=q^{n}d_{n}^{^{\prime }}c_{n}^{^{\prime }},\text{ \ }c_{n}^{^{\prime
}}b_{n}^{^{\prime }}=b_{n}^{^{\prime }}c_{n}^{^{\prime }}\text{ .}
\label{11}
\end{eqnarray}%
Following the same procedure as we obtain relations$(7)$ and with the limit
identity

\begin{equation}
\lim_{q\rightarrow 1}\frac{q^{n}-1}{q-1}=n\text{ },  \label{12}
\end{equation}%
we can have the following commutation relations of the $GL_{nh}(2)$ given by

\begin{eqnarray}
\left[ a_{n},c_{n}\right] &=&nhc_{n}^{2},\text{ \ \ \ \ \ \ \ \ \ \ \ \ \ }%
\left[ d_{n},b_{n}\right] =nh(D_{nh}-d_{n}^{2}),\text{ \ }  \notag \\
\left[ a_{n},d_{n}\right] &=&nhd_{n}c_{n}-nha_{n}c_{n},\text{ \ \ }\left[
d_{n},c_{n}\right] =nhc_{n}^{2},\text{ \ \ }  \label{13} \\
\left[ b_{n},c_{n}\right] &=&nha_{n}c_{n}+nhc_{n}d_{n},\text{ \ \ }\left[
b_{n},a_{n}\right] =nh(a_{n}^{2}-D_{nh})  \notag
\end{eqnarray}%
where

\begin{equation}
D_{nh}=\det_{h}M^{n}\equiv a_{n}d_{n}-c_{n}b_{n}-nhc_{n}d_{n},  \label{14}
\end{equation}%
which mean that $M^{n}$ does belong to the $h$-deformed quantum group $%
GL_{nh}(2)$.

In fact, by using the $R$-matrix for $q$-deformation\cite{6}:

\begin{equation}
R^{^{\prime }}=\left( 
\begin{array}{cccc}
1 & 0 & 0 & 0 \\ 
0 & q & 1-q^{2} & 0 \\ 
0 & 0 & q & 0 \\ 
0 & 0 & 0 & 1%
\end{array}%
\right) \text{ ,}  \label{15}
\end{equation}%
one can obtain the commutation relations(2) between the elements of $q$%
-deformed matrix $M^{^{\prime }}$ through the following equation

\begin{equation}
R^{^{\prime }}M_{1}^{^{\prime }}M_{2}^{^{\prime }}=M_{2}^{^{\prime
}}M_{1}^{^{\prime }}R^{^{\prime }}\text{ ,}  \label{16}
\end{equation}%
where $M_{1}^{^{\prime }}$, $M_{2}^{^{\prime }}$ are the graded matrices
with the following matrix elements respectively

\begin{eqnarray}
\left( M_{1}^{^{\prime }}\right) _{cd}^{ab} &=&\left( M^{^{\prime }}\otimes
I\right) _{cd}^{ab}=M_{c}^{^{\prime }a}\delta _{d}^{b},  \label{17} \\
\left( M_{2}^{^{\prime }}\right) _{cd}^{ab} &=&\left( I\otimes M^{^{\prime
}}\right) _{cd}^{ab}=M_{d}^{^{\prime }b}\delta _{c}^{a}\text{ .}  \notag
\end{eqnarray}%
And it is clear that the transformation between the deformed quantum groups $%
M^{^{\prime }}=gMg^{-1}$ leads to the similarity transformation between the
corresponding $R$-matrices for different quantum groups:

\begin{equation}
R=\left( g\otimes g\right) ^{-1}R^{^{\prime }}\left( g\otimes g\right) \text{
.}  \label{18}
\end{equation}%
Here product $\otimes $ acts in the same manner as $M_{1}^{^{\prime }}$ and $%
M_{2}^{^{\prime }}$ expressions(17). Through such transformation the $R$%
-matrix for the $h$-deformation can be obtained as the following

\begin{equation}
R_{h}\equiv \lim_{q\rightarrow 1}R=\left( 
\begin{array}{cccc}
1 & -h & h & h^{2} \\ 
0 & 1 & 0 & -h \\ 
0 & 0 & 1 & h \\ 
0 & 0 & 0 & 1%
\end{array}%
\right) \text{ .}  \label{19}
\end{equation}%
Applying Eq.$(4)$ and Eq.$(18)$, one can obtain the following equation in
the limit of $q\rightarrow 1$

\begin{equation}
R_{h}M_{1}M_{2}^{^{{}}}=M_{2}^{^{{}}}M_{1}R_{h}\text{ .}  \label{20}
\end{equation}%
So the $h$-commutation relations$(7)$ can be derived from the above equation
again. As a common sense, the $R$-matrix for $q$-deformation in the quantum
group $GL_{q^{n}}(2)$ is given by

\begin{equation}
R_{n}^{^{\prime }}=\left( 
\begin{array}{cccc}
1 & 0 & 0 & 0 \\ 
0 & q^{n} & 1-q^{2n} & 0 \\ 
0 & 0 & q^{n} & 0 \\ 
0 & 0 & 0 & 1%
\end{array}%
\right) \text{ .}  \label{21}
\end{equation}%
Because of the transformation$(9)$, we also have

\begin{equation}
R_{n}=\left( g\otimes g\right) ^{-1}R_{n}^{^{\prime }}\left( g\otimes
g\right) \text{ .}  \label{22}
\end{equation}%
So with the limit identity$(12)$, we can obtain the $R$-matrix for $h$%
-deformation in the quantum group $GL_{nh}(2)$ following by

\begin{equation}
R_{nh}\equiv \lim_{q\rightarrow 1}R_{n}=\left( 
\begin{array}{cccc}
1 & -nh & nh & \left( nh\right) ^{2} \\ 
0 & 1 & 0 & -nh \\ 
0 & 0 & 1 & nh \\ 
0 & 0 & 0 & 1%
\end{array}%
\right) \text{ .}  \label{23}
\end{equation}%
Also from Eq.$(9)$ and Eq.$(22)$, we derive the following equation in the
limit of $q\rightarrow 1$

\begin{equation}
R_{nh}M_{1}^{n}M_{2}^{n}=M_{2}^{n}M_{1}^{n}R_{nh}  \label{24}
\end{equation}%
and the commutation relations Eq.(13) of the quantum group $GL_{nh}(2)$ are
obtained through this equation and it shows that $M^{n}$ definitely belongs
to $GL_{nh}(2)$.

Next step, we will extend our discussion to quantum supergroup. We denote
supermatrix quantity as $\widetilde{Q}$ with $Q$ representing the quantity
for the common deformed group. Returning to the Manin quantum superplane
satisfying the relation $x\theta =q\theta x,$ $\theta ^{2}=0$, there exists
the $R$-matrix for the quantum supergroup $GL_{q}(1\mid 1)$ acting on the
superplane\cite{9}, where

\begin{equation}
\widetilde{R}^{^{\prime }}=\left( 
\begin{array}{cccc}
q & 0 & 0 & 0 \\ 
0 & 1 & 0 & 0 \\ 
0 & q-q^{-1} & 1 & 0 \\ 
0 & 0 & 0 & q^{-1}%
\end{array}%
\right) \text{ .}  \label{25}
\end{equation}%
The $h$-deformed quantum superplane satisfying the relation $x\theta =\theta
x+hx^{2},$ $\theta ^{2}=-h\theta x$ can also be obtained through the
singular transformation in the limit $q\rightarrow 1$. Conventionally, here
we take the transfer matrix $g$ as $\left( 
\begin{array}{cc}
1 & 0 \\ 
\frac{h}{q-1} & 1%
\end{array}%
\right) $. Observing that for $h$-deformed quantum supergroup the
deformation parameter must be a dual odd (Grassmann) number satisfying \ $%
h^{2}=0\cite{10}$, one can obtain the corresponding $R$-matrix for the
quantum supergroup $GL_{h}(1\mid 1)$ acting on the $h$-deformed quantum
superplane through the similarity transformation in Eq.$\left( 18\right) $:%
\begin{equation}
\widetilde{R}_{h}\equiv \lim_{q\rightarrow 1}\widetilde{R}=\left( 
\begin{array}{cccc}
1 & 0 & 0 & 0 \\ 
-h & 1 & 0 & 0 \\ 
h & 0 & 1 & 0 \\ 
0 & h & h & 1%
\end{array}%
\right) \text{ .}  \label{26}
\end{equation}%
Here, the only difference is the grading product $\otimes $. For quantum
supergroup, the product acts in the following manner

\begin{eqnarray}
\left( \widetilde{M}_{1}^{^{\prime }}\right) _{cd}^{ab} &=&\left( \widetilde{%
M}^{^{\prime }}\otimes I\right) _{cd}^{ab}=\left( -1\right) _{{}}^{c(b+d)}%
\widetilde{M}_{c}^{^{\prime }a}\delta _{d}^{b},  \notag \\
\left( \widetilde{M}_{2}^{^{\prime }}\right) _{cd}^{ab} &=&\left( I\otimes 
\widetilde{M}^{^{\prime }}\right) _{cd}^{ab}=\left( -1\right) ^{a(b+d)}%
\widetilde{M}_{d}^{^{\prime }b}\delta _{c}^{a}\text{ .}  \label{27}
\end{eqnarray}%
From the same relation as Eq.$\left( 20\right) $, one can obtain the
commutation relations of quantum supergroup $GL_{h}(1\mid 1)$, which is
given by

\begin{eqnarray}
ab &=&ba,\text{ \ \ \ \ \ \ \ \ \ \ \ \ \ \ \ \ \ \ }ac=ca+h(a^{2}+cb-da) 
\notag \\
bd &=&db,\text{ \ \ \ \ \ \ \ \ \ \ \ \ \ \ \ \ \ \ }cd=dc+h(d^{2}-cb-da)
\label{28} \\
b^{2} &=&0,\text{ \ \ \ \ \ \ \ \ \ \ \ \ \ \ \ \ \ \ \ }c^{2}=h(dc-ca) 
\notag \\
bc &=&-cb-h(ba-db),\text{ \ \ \ \ \ }ad=da+h(ba-db)\text{ .}  \notag
\end{eqnarray}%
The $n$-th power of supermatrix $\widetilde{M}^{^{\prime }}$ belongs to
quantum supergroup $GL_{q^{n}}(1\mid 1)$ and $R$-matrix for $%
GL_{q^{n}}(1\mid 1)$ is given by

\begin{equation}
\widetilde{R}_{n}^{^{\prime }}=\left( 
\begin{array}{cccc}
q^{n} & 0 & 0 & 0 \\ 
0 & 1 & 0 & 0 \\ 
0 & q^{n}-q^{-n} & 1 & 0 \\ 
0 & 0 & 0 & q^{-n}%
\end{array}%
\right) \text{ .}  \label{29}
\end{equation}%
So through the transformation like Eq.(22) and with the limit of $%
q\rightarrow 1$, we can have the $R$-matrix for quantum supergroup $%
GL_{nh}(1\mid 1)$ as following

\begin{equation}
\widetilde{R}_{nh}\equiv \lim_{q\rightarrow 1}\widetilde{R}=\left( 
\begin{array}{cccc}
1 & 0 & 0 & 0 \\ 
-nh & 1 & 0 & 0 \\ 
nh & 0 & 1 & 0 \\ 
0 & nh & nh & 1%
\end{array}%
\right)  \label{30}
\end{equation}%
Also following the same procedure as we obtain Eq.(24) we have the relation
for the $n$-th power of quantum supermatrix $\widetilde{M}^{n}$:

\begin{equation}
\widetilde{R}_{nh}\widetilde{M}_{1}^{n}\widetilde{M}_{2}^{n}=\widetilde{M}%
_{2}^{n}\widetilde{M}_{1}^{n}\widetilde{R}_{nh}\text{ ,}  \label{31}
\end{equation}%
which shows that the entries of the transformed quantum matrix $\widetilde{M}%
^{n}$ fulfil the commutation relations of the quantum supergroup $%
GL_{nh}(1\mid 1)$ as following%
\begin{eqnarray}
ab &=&ba,\text{ \ \ \ \ \ \ \ \ \ \ \ \ \ \ \ \ \ \ }ac=ca+nh(a^{2}+cb-da) 
\notag \\
bd &=&db,\text{ \ \ \ \ \ \ \ \ \ \ \ \ \ \ \ \ \ \ }cd=dc+nh(d^{2}-cb-da)
\label{32} \\
b^{2} &=&0,\text{ \ \ \ \ \ \ \ \ \ \ \ \ \ \ \ \ \ \ \ }c^{2}=nh(dc-ca) 
\notag \\
bc &=&-cb-nh(ba-db),\text{ \ \ \ \ \ }ad=da+nh(ba-db)\text{ .}  \notag
\end{eqnarray}%
Hence, we conclude that $\widetilde{M}^{n}$ belongs to $GL_{nh}(1\mid 1)$.

In summary, we show that the $n$-th power of $h$-deformed quantum matrix is
quantum matrix with the deformation parameter $nh$ and the same is also true
for the $n$-th power of $h$-deformed quantum supermatrix.

\end{document}